\documentclass[1p]{elsarticle}

\usepackage{hyperref}
\def\nn{\nonumber\\}
\def\bea{\begin{eqnarray}}
\def\eea{\end{eqnarray}}
\def\wt{\widetilde}
    
\def\Vec#1{\mbox{\boldmath $#1$}}

\usepackage{sidecap}  
\sidecaptionvpos{figure}{t}

\begin{document}

\begin{frontmatter}

\title{A mechanical model for diversified insect wing margin shapes}

\author{Yukitaka Ishimoto}
\address{Department of Machine Intelligence and Systems Engineering, Akita Prefectural University, Akita 015-0055, Japan}
\cortext[mycorrespondingauthor]{Corresponding author}
\ead{ishimoto@akita-pu.ac.jp}

\author{Kaoru Sugimura}
\address{Institute for Integrated Cell-Material Sciences (WPI-iCeMS), Kyoto University, Kyoto 606-8501, Japan}
\address{JST PRESTO, Tokyo 102-0075, Japan}

\begin{abstract}
The wings in different insect species are morphologically distinct with regards to
their size, outer contour (margin) shape, venation, and pigmentation.
The basis of the diversity of wing margin shapes remains unknown,
despite the fact that gene networks governing the \textit{Drosophila} wing development have been well characterised.
Among the different types of wing margin shapes, smoothly curved contour
is the most frequently found and implies the existence of a highly organised, multicellular mechanical structure.
Here, we developed a mechanical model for diversified insect wing margin shapes,
in which non-uniform bending stiffness of the wing margin is considered.
We showed that a variety of spatial distribution of the bending stiffness could reproduce diverse wing margin shapes.
Moreover, the inference of the distribution of the bending stiffness from experimental images indicates a common spatial profile among insects tested.
We further studied the effect of the intrinsic tension of the wing blade on the margin shape and on the inferred bending stiffness.
Finally, we implemented the bending stiffness of the wing margin in the cell vertex model of the wing blade,
and confirmed that the hybrid model retains the essential feature of the margin model.
We propose that in addition to morphogenetic processes in the wing blade,
the spatial profile of the bending stiffness in the wing margin can play a pivotal role in shaping insect wings.
\end{abstract}

\begin{keyword}
Mechanics, Morphogenesis, Insect, Wing
\end{keyword}

\end{frontmatter}

\begin{figure}[bt]
  \centering
  \resizebox{1.0\textwidth}{!}{
  \includegraphics[bb=0 0 409 168]{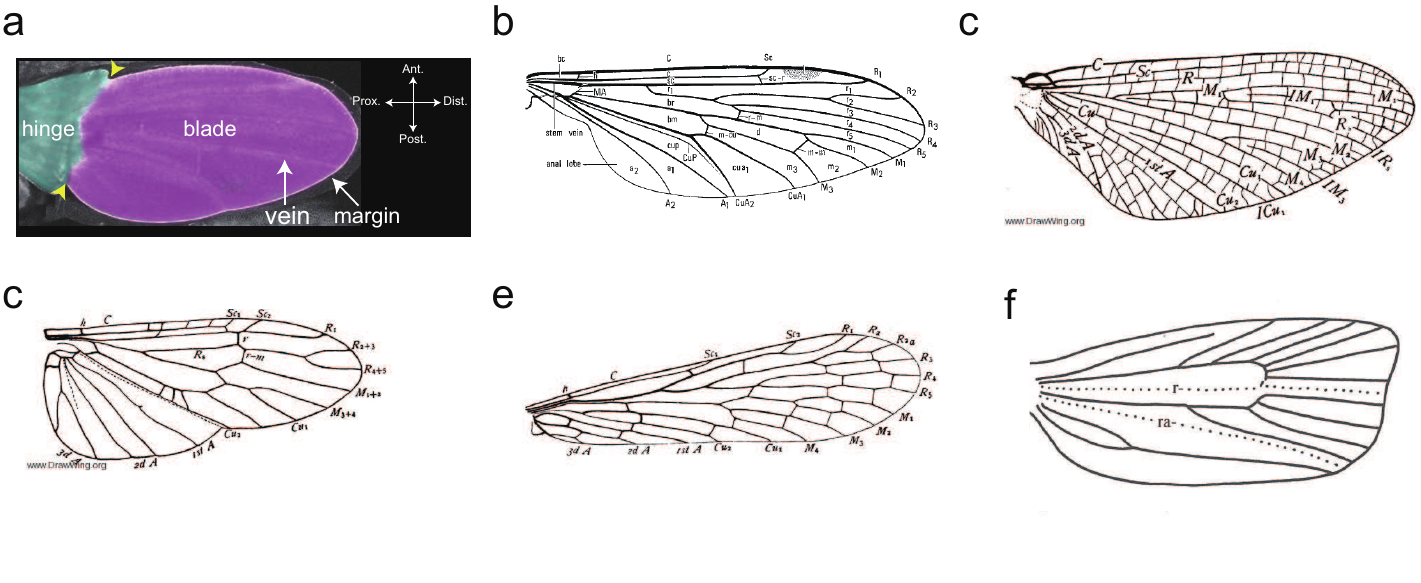}
  }
  \caption{Diverse wing margin shapes among insects. 
  (a) Image of \textit{Drosophila} pupal wing at 32 hr APF (after puparium formation).
  The endogenous DECadherin is labeled with GFP \cite{Huang2009}. 
  Yellow arrowheads indicate the positions of the hinge/blade/margin crossing points where the pinching forces act.
  The vertical and horizontal directions are aligned with the anterior-posterior and proximal-distal axes, respectively. 
  (b--f) Drawings of the wings of different insects.
  (b) Diptera (adapted from \cite{McAlpine1981}).
  (c) Heptageniidae (adapted from \cite{Comstock1918}).
  (d) Chloroperla (adapted from \cite{Comstock1918}).
  (e) Panorpa (adapted from \cite{Comstock1918}).
  (f) Tortricidae (adapted from \cite{Grodnitsky1962}).
  }
  \label{fig:realW}
\end{figure} 

\section{Introduction}
Insects have acquired the ability to fly with special appendages, \emph{i.e.} wings, 
which confer upon them adaptive fitness \cite{Grodnitsky1962, Dudley2002}.
Insect wing consists of the blade, vein, margin, sensory organs, and trachea.
The base of the wing is connected to the body via the hinge. 
Insect adult structures, including wings, develop from imaginal discs,
which arise as epithelial folds in the embryonic ectoderm and grow inside the larval body \cite{Held2002}.
In \textit{Drosophila}\footnote{
The insect wing development has been mostly studied using \textit{D. melanogaster} as a model organism.
A mechanical model formulated in this study corresponds to developmental events
at 15 to 27 hr APF (after puparium formation) in \textit{D. melanogaster.}
Nomenclature and spatiotemporal dynamics of the wing development are different among insects.
For instance, in many insects, the wing disc everts inside the larval body well before the puparium formation.},
upon evagination of the disc at the end of larval stages, the wing tissue forms a flat, epithelial bilayer (dorsal and ventral). 
Under the control of biochemical and mechanical interaction between cells,
the larval wing disc and pupal wing epithelium undergo extensive morphogenetic processes:
cells differentiate, grow, proliferate, move, and die to determine the final size, shape, and structure of the tissue [4--16].
Immediately after eclosion, epithelial cells die by programmed cell death and are absorbed into the thoracic cavity through the veins, 
leaving the exoskeleton \cite{Johnson1987, Kimura2004}.

The wings in different insect species are morphologically distinct with regards to their size, outer contour (margin) shape,
venation, and pigmentation (Fig. \ref{fig:realW}) [19--23].
For instance, dragonfly has a smoothly curved, elongated wing, whereas butterflies develop a fan-shaped wing.
Although gene networks governing the \textit{Drosophila} wing development have been well characterised,
very little work has been done on the basis of the diversity of wing margin shapes.
During evolution, organisms have tuned the unified mechanism of development, in particular gene network, to generate diverse morphologies [20--25].
One of the candidates for such unified basis of wing margin shape determination might be a highly organised, multicellular mechanical structure,
as suggested by the observation that many insect wings have a smoothly curved shape.

In this study, we formulate a mechanical model for simple yet diversified insect wing margin shapes.
We adopt the basic notion of Euler's elastica to wing development, where the stiff margin is
pinched by the hinge along the proximal boundary of the wing blade.
We then introduce non-uniform bending stiffness that depends on the position of the wing margin.
By using the model, we show that the spatial distribution of the bending stiffness could generate a variety of shapes
that resemble the smooth outer contours of natural insect wings.
We also infer the bending stiffness of the wing margin from experimental images. 
The inferred profiles of the bending stiffness of different insects are distinct, but all shared a common spatial domain structure.
These data imply that the conserved, mechanical machineries have been tuned to give rise to diverse wing margin shapes during evolution.

\section{Model}

\subsection{The biological basis for model formulation}
\label{sec:biology}
In this section, we explain why we focus on the mechanics of the wing margin.
Smoothly curved shapes of the wing margin, which can be found in different orders of insects (Fig. \ref{fig:realW}),
are reminiscent of a flexible, elastic rod under load such as largely deformed beam or semiflexible polymer chain 
including DNA, F-actin, and collagen [28--31].
In addition, studies reporting that the external force acting from the hinge stretches the wing
along the proximal-distal axis in \textit{Drosophila melanogaster} [10--12, 32]
are consistent with an idea that pinching forces act at the crossing points of the margin, hinge, and blade
(yellow arrowheads in Fig. \ref{fig:realW}(a)).
Because the crossing points are tied with the hinge, either the pinned or fixed boundary condition should be employed.
The observation that the angles of the margin at the crossing points relative to the hinge are not fixed  
during the wing development may suggest the pinned boundary condition,
whereas the attachment to the overlying cuticle via the Dumpy protein \cite{Etournay2015, Ray2015} may offer the fixed boundary condition.
Together, these suggest that an elastic, stiff margin is pinched by the extrinsic force and bends like the Euler's elastica
(Fig. \ref{fig:elastica}(a)) \cite{Euler}. 

As explained below, in our attempt to explain the morphological diversity of wings,
we consider the spatial distribution of the bending stiffness of the wing margin.
In the case of a beam, a mechanically uniform material,
the bending stiffness can be estimated by the Young's modulus multiplied by the second moment of area.
Actin cytoskeleton, molecular motors, extracellular matrix, and many other components would be involved in the regulation of
such a modulus of single cells and tissues [35-38]. 
Since the wing margin may not be mechanically uniform like the beam, 
the second moment of area should be changed according to cell morphogenetic processes 
such as cell proliferation and cell rearrangement \cite{Baonza2000, Takemura2011}.
Changing these biochemical, mechanical, and cellular parameters differentially affect the bending stiffness of tissue.
Although to our knowledge, non-uniform distribution of the bending stiffness of the wing margin at developmental stages has not been experimentally shown, 
the fact that the wing margin has three (\emph{i.e.} proximal anterior, distal anterior, and posterior) domains,
each of which contains a unique set of differentiated cells in \textit{D. melanogaster} \cite{Garcia-Bellido1972, Couso1994}, 
implies a spatially patterned mechanical structure.
Indeed, similar spatial patterns of cell differentiation along the margin have been reported in other insects \cite{vanBreugel1980, Yoshida2011}.

When the extrinsic stretching force acts on the wing, the area of the wing blade is kept nearly constant \cite{Aigouy2010}. 
Below we will mention that our model can be applied to such a case.
Because a veinless mutant wing develops a largely normal morphology in \textit{D. melanogaster} \cite{deCelis2003},
we defer a study on the effect of wing veins to future work.

\subsection{The simplest elastic model of the wing margin shape -- Euler's elastica}
\label{sec:uniformK}
\begin{figure}[bt]
  \centering
  \resizebox{1.0\textwidth}{!}{
  \includegraphics[bb=0 0 421 173]{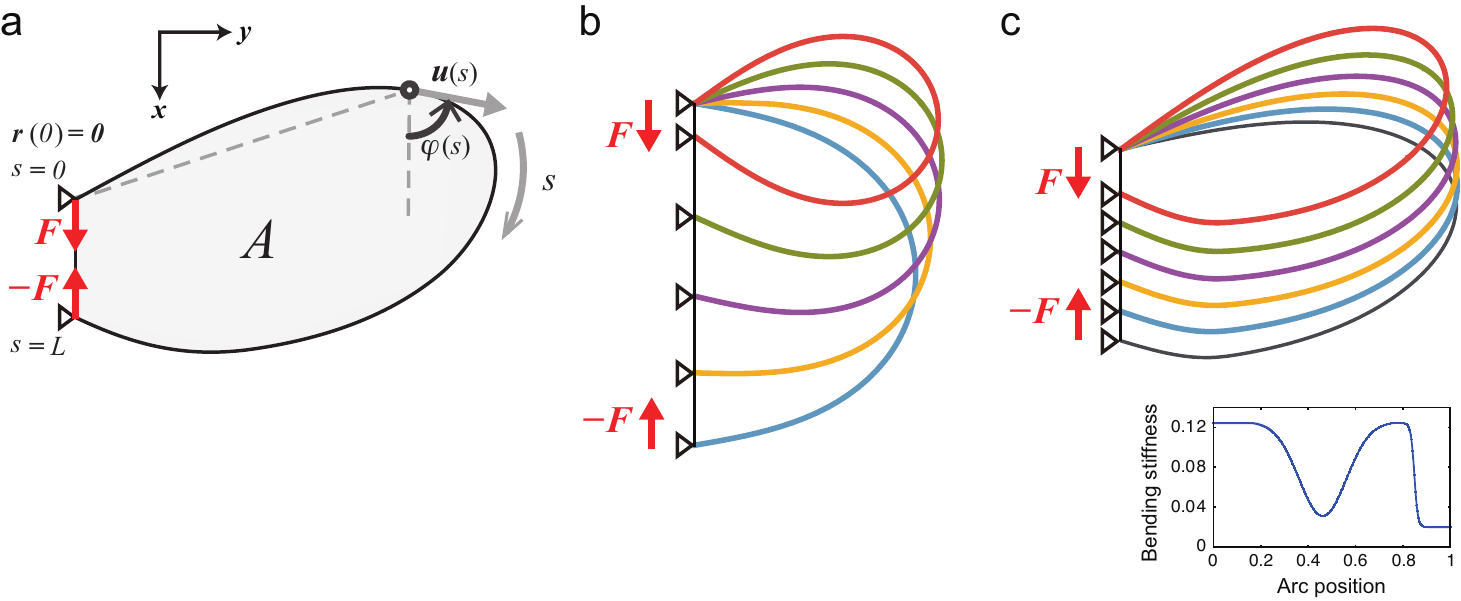}
  }
  \caption{The setting for the mathematical description of the elastica and the wing margin shapes, with two serial examples of the resulting curves.
  (a) The basic variables and parameters in the coordinate system for the elastica and the margin shapes. $s$ is the arc length parameter. $\varphi(s)$, $\Vec{u}(s)$, and $\Vec{r}(s)$ denote the tangential direction, the tangential vector, and the two-dimensional position of the margin segment at $s$, respectively. 
 $A$ is the area of wing blade. $\Vec{F}$ represents the pinching force acting at the ends, whose angle is set zero for convenience.
Triangles represent the pinned boundary condition.
  (b) A series of solutions of the Euler's elastica with the pinned boundary conditions for different $F$. The total arc length $L$ is constant.
  (c) A series of curves with a non-uniform bending stiffness $\kappa(s)$ as a set of solutions to Eq. (\ref{eq:nonuniformK}) for different $F$ with the pinned conditions. 
  $F$ values are different from those in (b). Note that the coordinate system is associated to the end position at $s=0$ so that it stays at the origin.
  The positional difference between the other ends in different simulations can be regarded as a consequence of such realignment of the positions.
  }
  \label{fig:elastica}
\end{figure} 

A naive and one of the simplest mechanical models of smooth margin shapes is the Euler's elastica:
Euler's mathematical and mechanical model of a thin strip or a thin rod and its resulting curves\footnote{
Note that Euler is not the founder of the problem, but the model has been historically called so \cite{history_elastica}.
}. 
Namely, a thin stiff rod bends or buckles in response to applied force, adapting a smooth, curved shape.
Mathematicians have idealised the rod and used two distinct formulations to solve the curves.
One formulates a force balance as equilibrium of moments along the rod, and the other formulates the bending energy and finds its minimum.
Both formulations give the same deterministic differential equation.
For a thin rod with length $L$ and bending stiffness $\kappa$, when the forces $\Vec{F}$ and $-\Vec{F}$ are acting on the two ends of the rod,
the equation can be written as
\bea
  \label{eq:Euler}
    \kappa \frac{d^2 \varphi(s) }{ds^2} = -F \sin \varphi(s), 
\eea
where $F \equiv|\Vec{F}|$, $s$ parameterises the position along the rod, {\it i.e.},
$s\in[0,L]$, and $\varphi(s)$ denotes the tangential angle of the rod at $s$ (Fig. \ref{fig:elastica}(a)). 
Eq. (\ref{eq:Euler}) is exactly the same as the equation of pendulum motion and can be solved analytically and exactly with the boundary conditions at the two ends 
(\ref{app:deriv}).
Corresponding to the initial conditions of the pendulum, such as being at rest or with some velocity at a given height, we have pinned and fixed boundary conditions. The pinned boundary condition means that the position is attached to its support by a freely rotating joint so that the angle is free. The fixed boundary condition means that the end point and its angle are totally fixed and supported as a beam is clamped to a wall. Mathematically speaking, the main difference between them is the value of $\frac{d\varphi(s)}{ds}$: it vanishes under the pinned condition, and it can be nonzero under the fixed boundary condition. In addition, there is the free boundary condition which means free to such constraints. In the case of an insect wing, the end points of the margin are attached to its support -- the hinge -- so either the pinned or fixed boundary condition is implied.
With the pinned conditions, the solutions of Eq. (\ref{eq:Euler}) for different $F$ are shown in Fig. \ref{fig:elastica}(b).

For later convenience, we further introduce other variables and parameters for the wing margin shape as shown in Fig. \ref{fig:elastica}(a). The tangential vector $\Vec{u}(s)$ along the margin and the two-dimensional position $\Vec{r}(s)$ of the margin segment at $s$ are 
\bea
  \label{def:u and r}
   \Vec{u}(s) &=& \left( \cos\varphi(s), \sin\varphi(s) \right)^T , 
\nn
   \Vec{r}(s) &=& \int_0^s dt\, \Vec{u}(t),
\eea
where the upper suffix $T$ stands for the transposition; we set $\Vec{r}(0)\equiv \mathbf{0}$ without loss of generality.
The area $A$ of the wing blade can then be given by
$  A = \frac12 \int_{0}^{L} ds \left( \Vec{u}(s) \times \Vec{r}(s) \right)$. 
Note that we can redefine $s$ in a unit segment, $s\in [0,1]$, and delete $F$ from the equation by nondimensionalisation (\ref{sec:nondimensionalization}).

In the next section, we develop a mechanical model with a non-uniform bending stiffness $\kappa$ along the concept of Euler's elastica.

\subsection{A mechanical model for insect wing margin shapes}
\label{sec:nonuniformK}
The deterministic equation with the non-uniform bending stiffness $\kappa(s)$ leads to 
\bea
  \label{eq:nonuniformK}
    \kappa(s) \frac{d^2 \varphi(s) }{ds^2} + \frac{d \kappa(s)}{ds} \frac{d \varphi(s) }{ds}= -F \sin \varphi(s), 
\eea
where $\kappa(s)$ becomes a distribution along $s$.
Its derivation is included in \ref{sec:derivation_extended}.
When $\kappa(s)$ is constant, the equation reduces to that of Euler's elastica.
In general, in the presence of the non-uniform distribution $\kappa(s)$, one cannot expect to obtain an exact, analytic solution.
Instead, we invoke a numerical calculation to find a wing margin shape for a given distribution of the bending stiffness.
We use the pinned condition, which is consistent with the system considered (Sec. \ref{sec:biology} and Sec. \ref{sec:uniformK}). 
By changing the magnitude of the external force $F$, 
or by changing the overall factor of $\kappa(s)$ in the nondimensionalised version of Eq. (\ref{eq:nonuniformK}),
the different margin shapes can be obtained (Fig. \ref{fig:elastica}(c) and Sec. \ref{sec:simulation}).

There are a few notes on computing the margin shape by using our model.
First, the external forces and moments must be balanced among themselves; otherwise, the wing moves and rotates by the external residual force and torque.
The forces are cancelled by definition while the torques must satisfy
\bea
  \label{eq:torque}
  \left( \Vec{r}(L) - \Vec{r}(0) \right) \times \Vec{F} = 0.
\eea
Second, when the fixed boundary conditions are imposed such as $\frac{d}{ds}\varphi(0) \ne 0$, one must be aware of another natural constraint,
which in the case of Euler's elastica corresponds to the energy conservation law of the pendulum.

By the virtue of the nondimensionalisation in \ref{sec:nondimensionalization},
one can apply this model to the case with a fixed area of the wing blade, which has been reported in \textit{D. melanogaster} [10]:
calculate the wing shapes and areas for a common $\kappa(s)$ with different multiplicities in the nondimensionalised equation and then rescale or dimensionalise the quantities such as $\kappa(s)$, $F$, and $L$ so that the wing blade area is kept constant. 
This can be directly implemented in the current modelling framework.

\subsection{The homogeneous pressure model}
\label{sec:extended}
\begin{figure}[bt] 
  \centering
  \resizebox{0.6\textwidth}{!}{
  \includegraphics[bb=0 0 321 110]{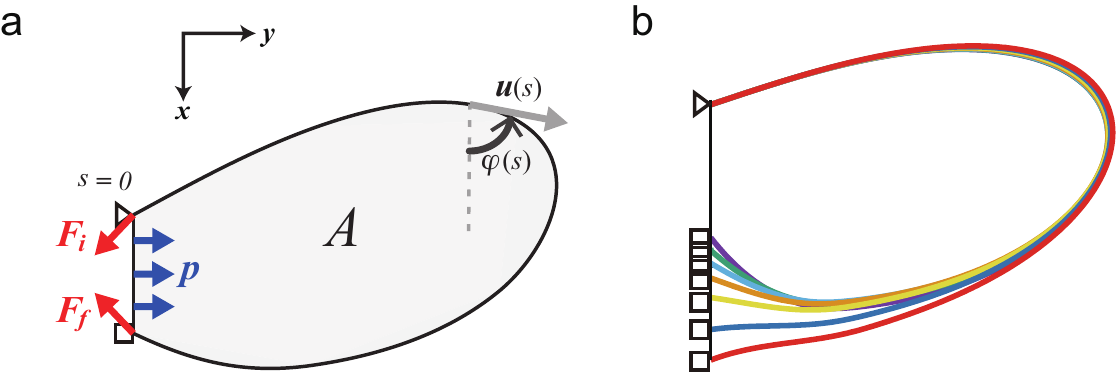}
  }
  \caption{ A schematic of the applied forces and the resulting curves in the homogeneous pressure model.
  (a) The applied forces and the basic settings for the homogeneous pressure model.
  The internal pressure/tension $p$ balances with the forces $\Vec{F}_i$ and $\Vec{F}_f$ acting on the margin ends.
  Triangle and square represent the pinned and fixed boundary condition, respectively.
  (b) A series of curves for different $p$ in the homogeneous pressure model. 
  The boundary conditions are pinned at $s=0$ and fixed at $s=L$.
  The non-uniform bending stiffness $\kappa(s)$ in Fig. \ref{fig:elastica}(c) is used. Apparent changes in positions of the end at $s=L$ are due to the coordinate system being fixed to the end at $s=0$. }
  \label{fig:pressuremodel}
\end{figure} 

Motivated by the observation that \textit{Drosophila} pupal wing develops tensile tissue stress upon tissue stretching by some external force [10--12, 32], 
we next implement the homogeneous tension of the wing blade as additional mechanical ingredient of the model (Fig. \ref{fig:pressuremodel}(a)).
We call such a derivative of our model the homogeneous pressure model in what follows.
Given an internal homogeneous pressure $p$ of the wing blade, which is negative in the case of tensile force, the equation can be derived as 
\bea
  \label{eq:p}
  \frac{d}{ds} \left( \kappa(s) \frac{d \varphi(s)}{ds} \right)
   = -F \sin\varphi(s) + p \Vec{u}(s) \cdot \left( \Vec{r}(s) - \frac{\Vec{r}(L)}{2} \right).
\eea
The homogeneous hydrostatic pressure $p$ appears in the second term of the right hand side. 
Note that when we deal with the energy of the wing while keeping the pressure $p$ constant,
the internal area $A$, which is mechanically conjugate to $p$, is not a conserved quantity.
Its detailed derivation is presented in \ref{sec:derivation_extended}.
Another important note regards the balance equation of the external forces. 
In the system considered, the tensile force is acting through the proximal boundary of the wing;
thus the tensile force must be balanced with the non-parallel forces $\Vec{F}_{i}$ and $\Vec{F}_{f}$ acting on the ends of the margin. 
Introducing a unit vector $\Vec{n}$ normal to the vector $\Vec{r}(L)-\Vec{r}(0)$ pointing inwards to the wing, this condition can be expressed as:
\bea
    p |\Vec{r}(L)-\Vec{r}(0)| \Vec{n} + \Vec{F}_i + \Vec{F}_f = 0.
\eea
The difference between this and afore-mentioned force balance equations is only the forces that are normal to $\Vec{r}(L)-\Vec{r}(0)$.
The components of $\Vec{F}_i$ and $\Vec{F}_f$ parallel to the proximal boundary are the same as before, $\Vec{F}$ and $-\Vec{F}$, and $F$ is given by $|\Vec{F}|$.
Therefore,
Eq. (\ref{eq:torque}) is automatically satisfied.

By changing the internal tension, or the negative pressure $p$, for fixed $\kappa(s)$,
one can draw a variety of curves depicted in Fig. \ref{fig:pressuremodel}(b).
The boundary conditions are the pinned condition at $s=0$ and the fixed condition at $s=L$.
Such asymmetry in the boundary conditions may be realised 
by differential attachment strength to the cuticle via the Dumpy protein \cite{Etournay2015, Ray2015}.  
Indeed, Dumpy accumulates more at the anterior end than at the posterior end (Fig. 2H of \cite{Etournay2015}), 
which may correspond to the fixed boundary condition at the anterior end and the pinned boundary condition at the posterior end. 
Note that the choice of pinned/fixed boundary conditions is opposite in our simulation due to a technical reason.

\section{Numerical results}
\label{sec:result}

\begin{figure}[bt] 
  \centering
  \includegraphics[width=11cm,bb=0 0 428 232]{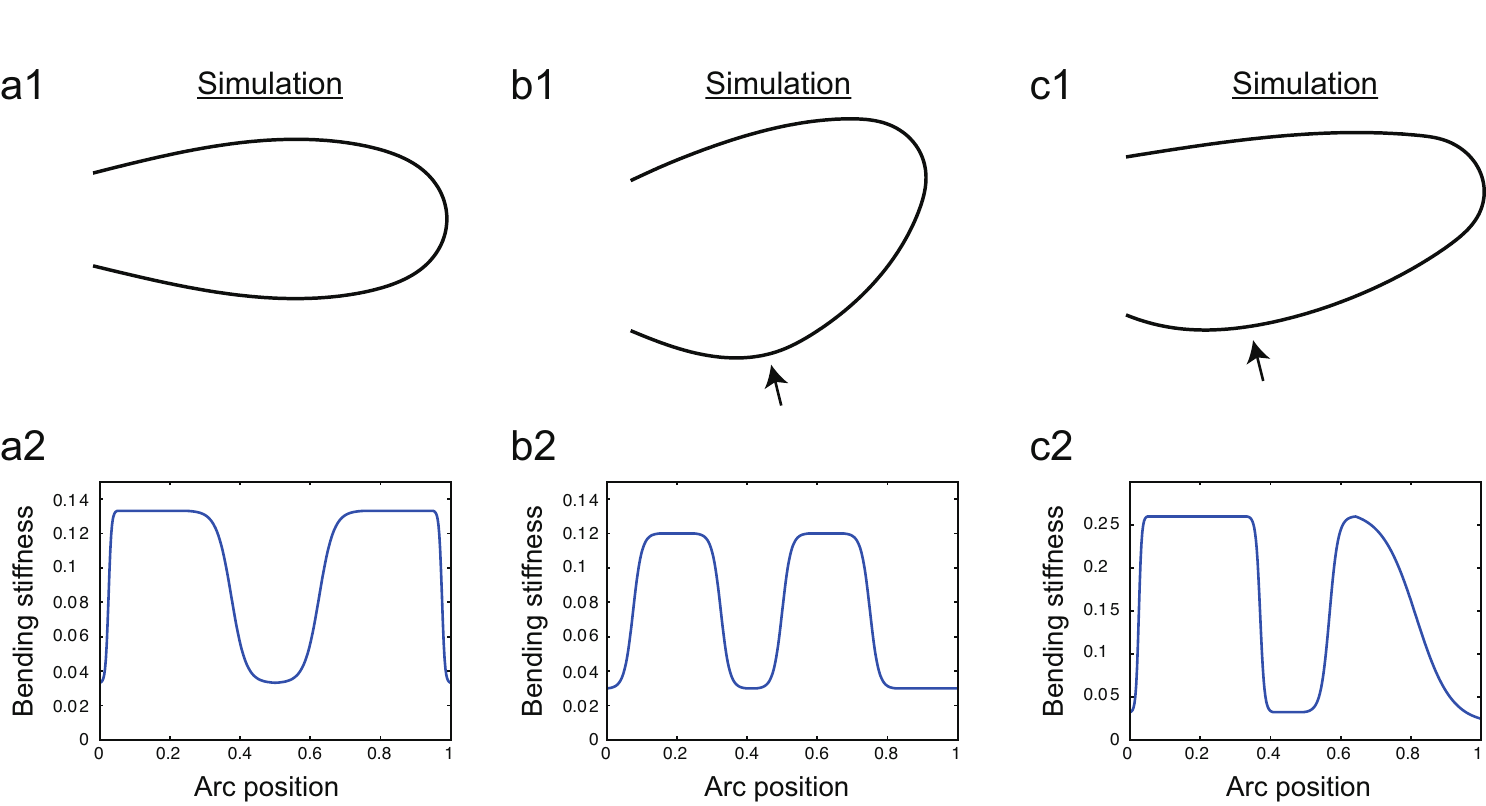}
  \caption{A simple case study showing the relationship between the simulated wing margin shape and the distribution of $\kappa(s)$.
   (a1, b1, c1) Curves obtained by simulation. Arrows point to the bulge in the posterior region (b1, c1).
   (a2, b2, c2) The corresponding non-uniform bending stiffness $\kappa(s)$. The parameters are normalised so that $F$ is included in $\kappa(s)$.
   No pressure term is considered and the pinned-pinned boundary condition is used.
   }
  \label{fig:nonuniformK-Simple}
\end{figure}  

\begin{figure}[bt] 
  \centering
  \includegraphics[width=10cm,bb=0 0 475 409]{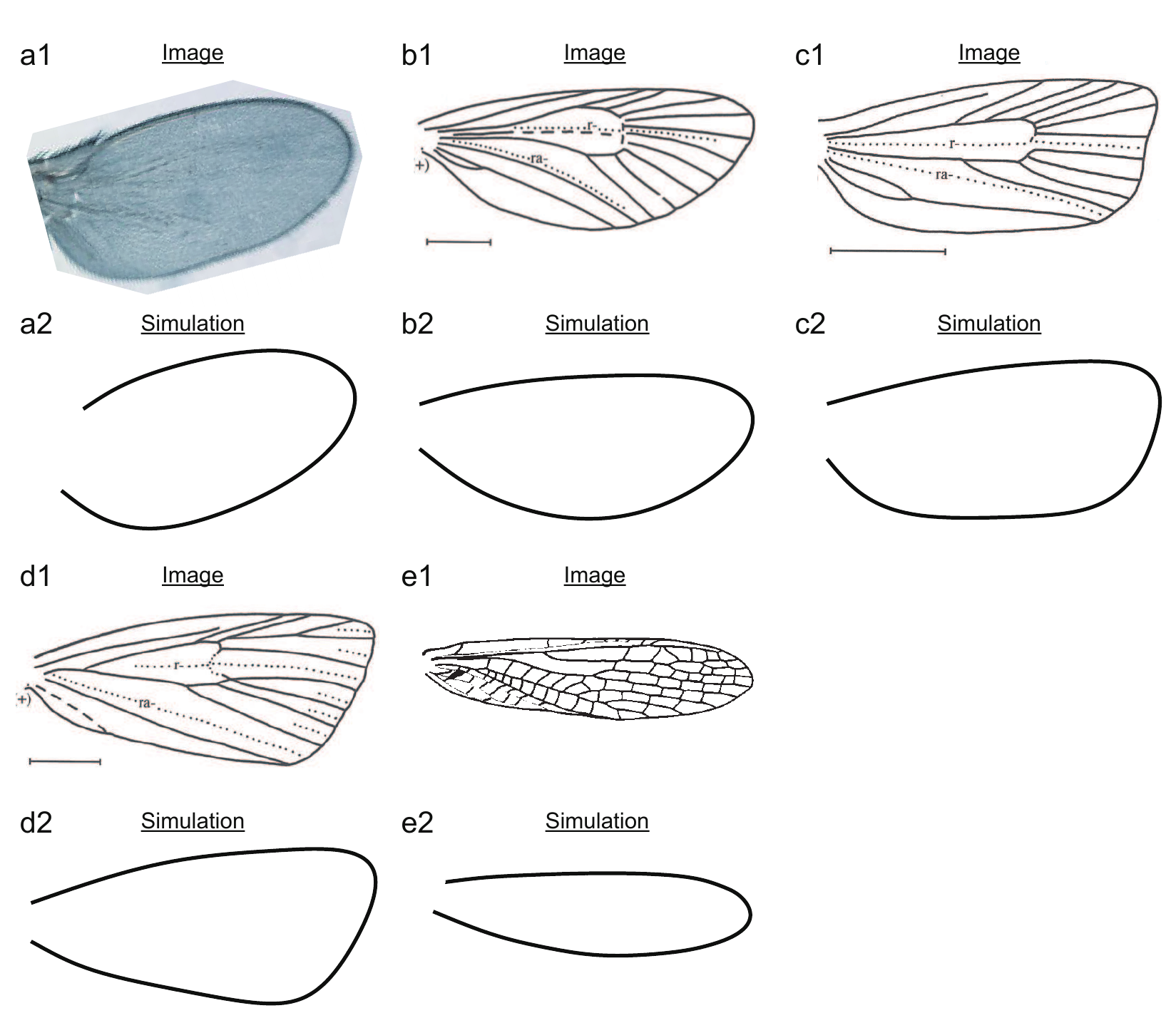}
  \caption{Simulated diversified wing margin shapes. 
  (a1--e1) A photograph (a1) and drawings (b1-e1) of wings of different insects.
  A veinless mutant of \textit{D. melanogaster} (adapted from \cite{deCelis2003}) (a1),
  Zygaenidae (adapted from \cite{Grodnitsky1962}) (b1), 
  Tortricidae (adapted from \cite{Grodnitsky1962}) (c1), 
  Crambidae (adapted from \cite{Grodnitsky1962}) (d1), 
  Eustheniidae (adapted from \cite{Bethoux2005}) (e1).  
  (a2--e2) Simulated wing margin shapes with different distributions of the bending stiffness $\kappa(s)$. 
  In simulations, no pressure term is considered and the parameters are normalised as in Fig. \ref{fig:nonuniformK-Simple}. The pinned-pinned boundary condition is used.
   }
  \label{fig:nonuniformK}
\end{figure}  

\begin{figure}[bt] 
  \centering
  \includegraphics[width=8cm,bb=0 0 294 241]{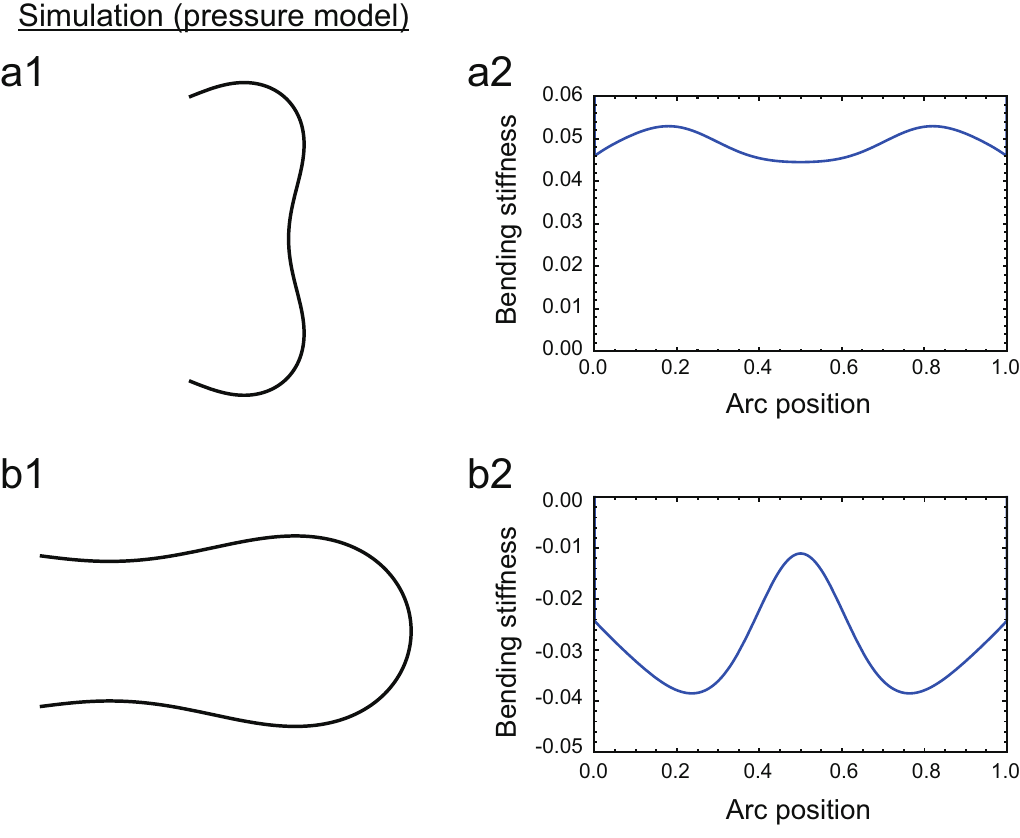}
  \caption{Typical buckling solutions of Eq. (\ref{eq:p}) for nonzero $p$ with the pinned boundary conditions.
  (a1) A buckled shape with tension, {\it i.e.} negative $p$. (b1) A buckled shape with positive pressure.
  (a2, b2) The corresponding non-uniform bending stiffness $\kappa(s)$. The parameters are normalised.}
  \label{fig:extended}
\end{figure} 

\subsection{Simulated wing margin shapes}
\label{sec:simulation}
Let us draw wing margin shapes by using our model for given distributions of the bending stiffness
(Fig. \ref{fig:nonuniformK-Simple} and Fig. \ref{fig:nonuniformK}).
We solve the nondimensionalised version of Eq. (\ref{eq:nonuniformK}) with the pinned condition at $s=0$ step by step using Euler's method.
The minimum mathematical requirement for $\kappa(s)$ is that it is at least once differentiable to fit to Eq. (\ref{eq:nonuniformK}).
Another physical requirement is that $\kappa(s)$ takes only a nonnegative value. 
Note that, if we abandon this physical requirement, one can create any smooth shape by unbounding and tuning $\kappa(s)$,
which is unrealistic in our current setup.
Since we do not know the initial condition of $\Vec{u}(0)$, {\it i.e.} $\varphi(0)$,
which is consistent with $r_y(0)=r_y(L=1)=0$ and the pinned condition at $s=1$,
its iterative search is also implemented in our simulations.

We first present a simple case study on a relationship between $\kappa(s)$ and the wing margin shape.
By using the symmetric distribution of $\kappa(s)$ along the AP axis (Fig. \ref{fig:nonuniformK-Simple}(a2)),
we obtain a shape mirrored about the AP axis (Fig. \ref{fig:nonuniformK-Simple}(a1)).
We found that the smaller value of $\kappa(s)$ in the distal region is necessary to obtain such a wing-like shape. 
By shifting the $\kappa(s)$ curve towards the anterior end (Fig. \ref{fig:nonuniformK-Simple}(b2)), the position of the distal tip of the wing follows 
(Fig. \ref{fig:nonuniformK-Simple}(b1)).  
Moreover, a bulge in the posterior region
is formed (arrow in Fig. \ref{fig:nonuniformK-Simple}(b1)),
which could not be produced by simulations in previous studies (\cite{Ray2015}).
The position of the bulge corresponds to a soft region posterior to the second peak of $\kappa(s)$.
In the first two examples, we have used a rather stepwise form of $\kappa(s)$. By giving a longer and smoother tail to the posterior peak of $\kappa(s)$ (Fig. \ref{fig:nonuniformK-Simple}(c2)),
the posterior bulge becomes smaller (arrow in Fig. \ref{fig:nonuniformK-Simple}(c1)),
and the resultant curve is reminiscent of a veinless mutant of \textit{D. melanogaster} \cite{deCelis2003}.
As a whole, this case study highlights characteristics of $\kappa(s)$ critical to shaping the wing margin.

We next attempt to create diverse margin shapes by elaborately constructing $\kappa(s)$.
Because even a simple distribution of the bending stiffness gives a shape reminiscent of an insect wing 
(Fig. \ref{fig:nonuniformK-Simple}), 
we manually assign some smoothly connected functional forms to $\kappa(s)$ as was performed in Fig. \ref{fig:nonuniformK-Simple},
referring to an inferred bending stiffness from the actual images of various insects.
Note that the former is the intuitive construction of the distribution by hand, and the latter is the mathematically derived inference of the distribution,
which will be given in Sec. \ref{sec:3.2-inference}.
The simulated margin shapes with manually constructed distributions of $\kappa(s)$
are clearly similar to those of real insects (Fig. \ref{fig:nonuniformK}).
The first example (Fig. \ref{fig:nonuniformK}(a1, a2)) shows an image of a veinless mutant of {\textit{D. melanogaster}
\cite{deCelis2003} and our simulation result.
For the simulation, we have used the asymmetric but similar curves of $\kappa(s)$ for the anterior and posterior regions, 
while a much softer, concave shape is assigned to the distal and posterior-end regions.
The following three sets of examples in Fig. \ref{fig:nonuniformK}(b1--d1, b2--d2) show the drawings of Zygaenidae, Tortricidae, and Crambidae
in \cite{Grodnitsky1962} and their corresponding simulation results.  
The results are reasonably good against the images,
although the details of $\kappa(s)$ had to be fine-tuned in the distal regions in the cases of Fig. \ref{fig:nonuniformK}(c2, d2). 
A slightly concave shape in the distal region of Fig. \ref{fig:nonuniformK}(c1) is outside the scope of the model without the internal pressure.
We will mention the reason in regards to the pressure shortly.
The last example shows an elongated shape of wings in Eustheniidae \cite{Bethoux2005}, which can be also found in dragonfly.
The wing margin shapes of those insects appear more or less anterior-posteriorly symmetric.
However, we have used significantly different values to $\kappa(s)$ of those regions, which will be inferred in the next section.
In our simulation, the shapes were sufficiently robust against small perturbations in $\kappa(s)$ in most part of the margin,
while some seemed sensitive to such perturbations, giving a potential source of the morphological diversity.
We speculate that the sensitivity comes from either characteristic values of curvature, or comparatively small values of the bending stiffness,
or the positions where the connection to the veins might be influential.
For the last possibility, we will discuss in Sec. \ref{sec:kappasignificance}.

By using the homogeneous pressure model with the internal pressure $p$, the wing margin can be buckled (Fig. \ref{fig:extended}).
Such buckled shapes cannot be obtained by using the original model simply because the original model describes a pinched form of a straight line whose curvature along $s$ should be positive everywhere when the pinned-pinned condition is imposed.

\subsection{Inference of the bending stiffness from experimental images and analysis on the relevance of the internal tension}
\label{sec:3.2-inference}
For a given distribution of $\kappa(s)$, from Eq. (\ref{eq:nonuniformK}), we have calculated the form of $\varphi(s)$ as the margin shape.
In reverse, for a given $\varphi(s)$, the distribution of $\kappa(s)$ can be calculated up to its initial value.
Therefore, schematically rewriting Eq. (\ref{eq:nonuniformK}) for $\kappa(s)$ as below and using the nondimensionalisation described in \ref{sec:nondimensionalization},
we can infer $\kappa(s)$ from an experimental image with a value of $\kappa(0)$:
\bea
  \dot \kappa(s) = - \frac{\kappa(s) \ddot\varphi(s) + \sin\varphi(s)}{\dot\varphi(s)} ,
\eea
where the dot ($\cdot$) stands for the $s$-derivative.

\begin{figure*}[bt] 
  \centering
  \includegraphics[width=13cm,bb=0 0 468 346]{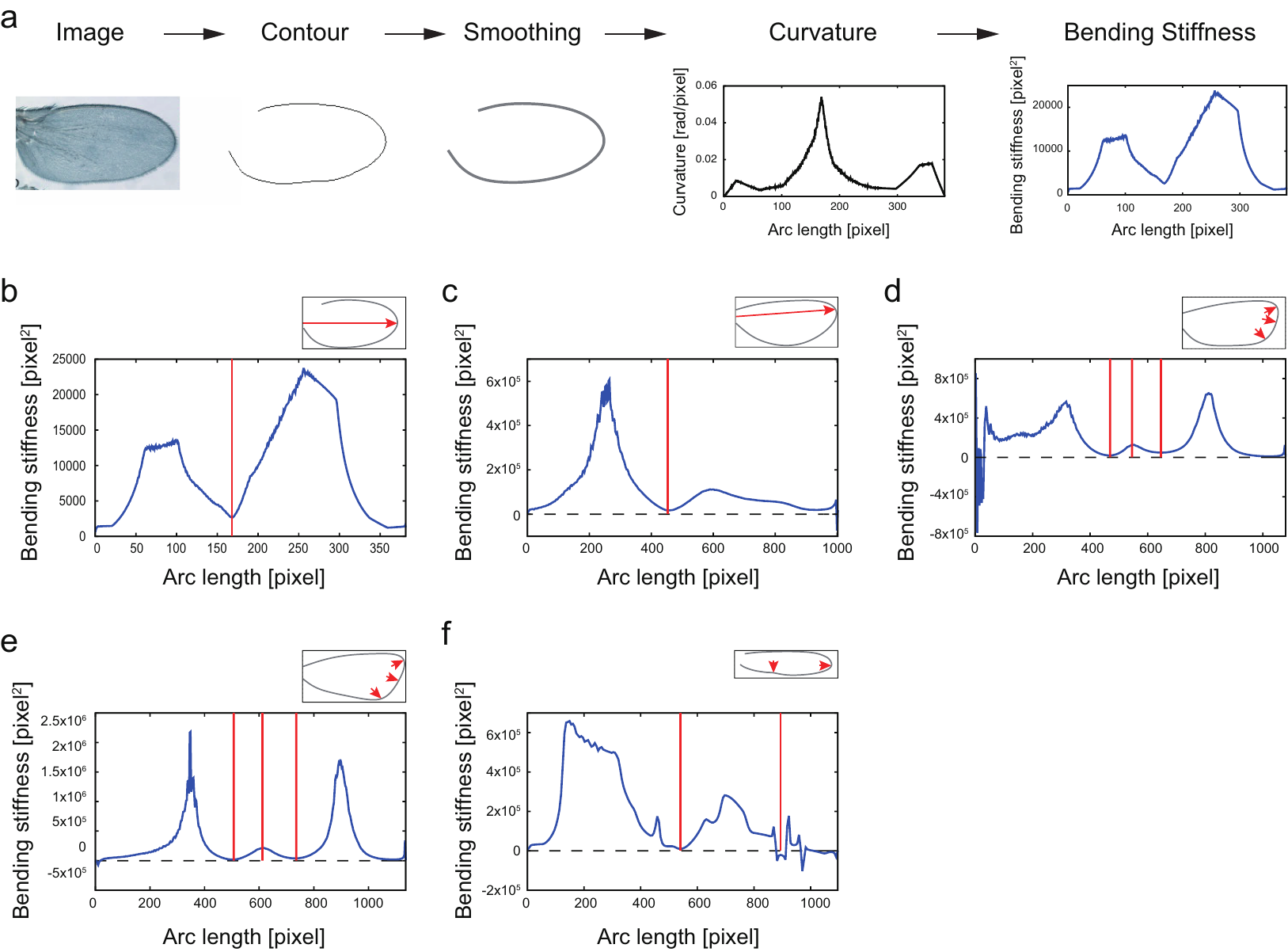}
  \caption{The inferred distributions of the bending stiffness $\kappa(s)$.
  The arc length parameter $s$ is given in the unit of pixel. The pinching force $F$ is normalised, and so is $\kappa(s)$ by $F$ as in \ref{sec:nondimensionalization}.
  Accordingly, $\kappa(s)$ is given in the unit of squared pixel.
  (a) A procedure of image processing for the inference of the bending stiffness $\kappa(s)$.
  A margin contour is extracted from a raw image, and the data set $\{\varphi(s),\dot\varphi(s),\ddot\varphi(s)\}$ is estimated from a generated smooth curve. 
  The bending stiffness is inferred through Eq. (\ref{eq:diff}).
  (b--f) A set of contour images and inferred bending stiffness. 
  Arrows in the contour images indicate the two-dimensional positions of the inserted red vertical lines in the plots of the inferred stiffness. 
  For the values of $\kappa(s)$ in the SI unit, the values in the plots should be rescaled by the total length of the margin and $F$ in the corresponding units. }  
  \label{fig:infer}  
\end{figure*} 

Let us translate the above differential equation into a difference equation and infer the value of $\kappa(s)$ from images of insect wings.
Eq. (\ref{eq:nonuniformK}) can be reduced to the difference equation
\bea
  \label{eq:diff}
  \dot \kappa(s) &=& -
  \frac{
    \left\{ \kappa(s-\Delta s) + \frac{\Delta s}2 \dot \kappa(s-\Delta s) \right\} 
      \ddot \varphi(s) + \sin\varphi(s) }
    { \dot \varphi(s) + \frac{\Delta s}2 \ddot \varphi(s) },
\nn
  \kappa(s) &=& \kappa(s-\Delta s)+\left(\dot\kappa(s-\Delta s)+\dot\kappa(s)\right) \frac{\Delta s}{2}
\eea
where $\Delta s$ is the segment length between nearest data points and $\Delta s \ll 1$ is assumed (refer to \ref{sec:inference} for its detailed derivation). 
The equation indicates that one can infer the values of $\left\{ \kappa(s), \dot\kappa(s) \right\}$ from a data set of $\{\varphi(s), \dot\varphi(s), \ddot\varphi(s)\}$ and ${\kappa(0), \dot\kappa(0)}$, step by step.
To obtain such data set, we first extract a margin contour from an experimental image (Fig. \ref{fig:infer}(a)). 
We then perform smoothing of the contour and resampling from the smoothed one, and calculate the curvature
from which the bending stiffness $\kappa(s)$ is inferred. 
The initial values are set by $\kappa(0)=1$ and $\dot\kappa(0)=0$ for simplicity.
By a simple reverse engineering, we can draw a very similar shape to the original wing image (data not shown).

From the images shown in Fig. \ref{fig:nonuniformK}(a1--e1),
we inferred the profiles of the bending stiffness $\kappa(s)$ (Fig. \ref{fig:infer}(b--f)).
Interestingly, in all cases, the bending stiffness exhibits asymmetric peak profiles in the anterior and posterior domains
and takes small values in the distal domain,
suggesting the existence of a conserved spatial structure among the insects tested
(discussed in Sec. \ref{sec:summary} and Sec. \ref{sec:kappasignificance}).

By further extending the difference equation in Eq. (\ref{eq:diff}) to the case with internal tension $p<0$, 
we can infer the bending stiffness in the presence of tension.
By varying the total pressure, $p |\Vec{r} (L)|$, from $0$ to $-0.2 F$, we found that
$\kappa(s)$ values in some regions fall below $0$ when $p |\Vec{r} (L)|$ is as large as $-0.01 F$ (Fig. \ref{fig:inferP}).
Therefore, only $p|\Vec{r}(L)|>-T F$ with $T\simeq 0.01$ is allowed in the examined case. 
This suggests that the internal tension is, at most, two orders of magnitude smaller than the pinching force at the ends of the wing margin
(discussed in Sec. \ref{sec:mathnotes}).

\begin{figure*}[bt] 
  \centering
  \includegraphics[width=9.5cm,bb=0 0 217 110]{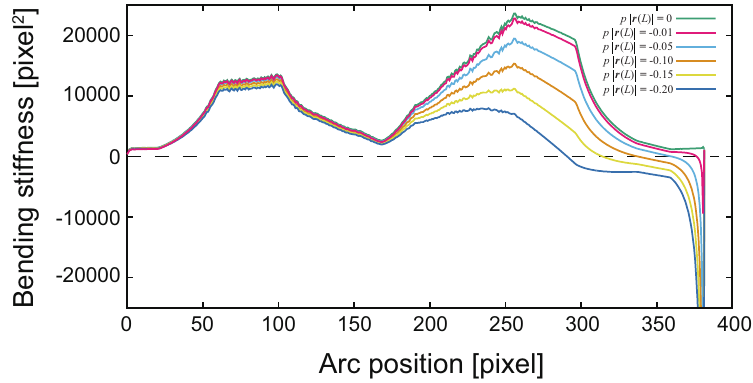}
  \caption{The inferred bending stiffness $\kappa(s)$ with different values of $p$ for the image shown in Fig. \ref{fig:nonuniformK}(a1).
  The units for the arc length parameter and $\kappa(s)$ are the same as in Fig. \ref{fig:infer}. 
  The pressure $p$ is given in the form of the total pressure $p |\Vec{r}(L)|$ in units of $F$. }
  \label{fig:inferP}
\end{figure*}

\section{An integrated mechanical model of the wing blade and margin}
\label{sec:IntegrateModel}

During wing development, cells in the wing blade and margin undergo extensive morphogenetic processes,
which are regulated by  biochemical and mechanical interactions between cells [4--16].
In this section, we address if the spatial profile of the bending stiffness of the wing margin can significantly 
contribute to the wing shape determination in the presence of other mechanical ingredients in the wing blade.
For this purpose, we formulate an integrated mechanical model of the wing margin and blade.
To simulate the \textit{Drosophila} pupal wing morphogenesis, 
the cell vertex model, which was formulated by Honda for modelling epithelial mechanics [46--49],
has been extensively used and has proven its validity for analysing morphogenetic cell processes, 
such as cell rearrangements, proliferations, cell shape changes due to applied forces, and so on
\cite{Aigouy2010, Sugimura2013, Ray2015}. 
We therefore construct a simple hybrid model of the stiff wing margin and the cell vertex model,
and perform simulations with different distributions of the bending stiffness of the wing margin.

The cell vertex model is a cell-based discrete model, so our mechanical model of the wing margin
should also be discretised in the same fashion to merge the models. Following convention, 
we use the energy formulation of the cell vertex model and the following energy function of the margin is added to it:
\bea
  \label{eq:bending_energy_on_vertex}
  E_{margin} &=& \sum_{i\in {\rm margin}} 2\, \kappa_i \sin^2 \frac{\theta_i}{2},
\eea
where $i$ stands for the vertex number along the margin, and $\kappa_i$ and $\theta_i$ are the bending stiffness and the angle associated to it.
The angle $\theta_i$ is defined by the angle difference between neighbouring orientation vectors 
analogous to $\Vec{u}(s)$ in Fig. \ref{fig:elastica}: the vector
connecting the $(i-1)$-th and $i$-th vertices, and that of the $i$-th and $(i+1)$-th vertices along the margin. 
In a sufficiently relaxed state of the cell vertex model,
each cell junction at the tissue boundary tends to take a similar length;
thus, we assign the values of the bending stiffness according to the margin vertex numbering counted from one end of the margin.
If the discretised input of the bending stiffness mismatches with the numbering, linear interpolation is applied.
To compare with the results in the preceding sections, we neglect the hinge region and 
make a straight boundary at the proximal side of the wing blade. 
Implementing the pinned boundary conditions on the two ends of the margin, 
we ran the simulations for, at least, a few times longer than the relaxation time of the cell vertex model. 
We used the symmetric and asymmetric distributions of the bending stiffness as in Fig. \ref{fig:nonuniformK-Simple}, 
and the uniform distribution of the bending stiffness to compare with. 
The results in Fig. \ref{fig:hybrid} show distinct shapes for different distributions of the bending stiffness.
This supports the idea that even in the presence of other mechanical processes in the wing blade, 
the stiffness of the margin may provide a key role in shaping the wing margin.

\begin{figure*}[bt] 
  \centering
  \includegraphics[width=13.5cm,bb=0 0 449 99]{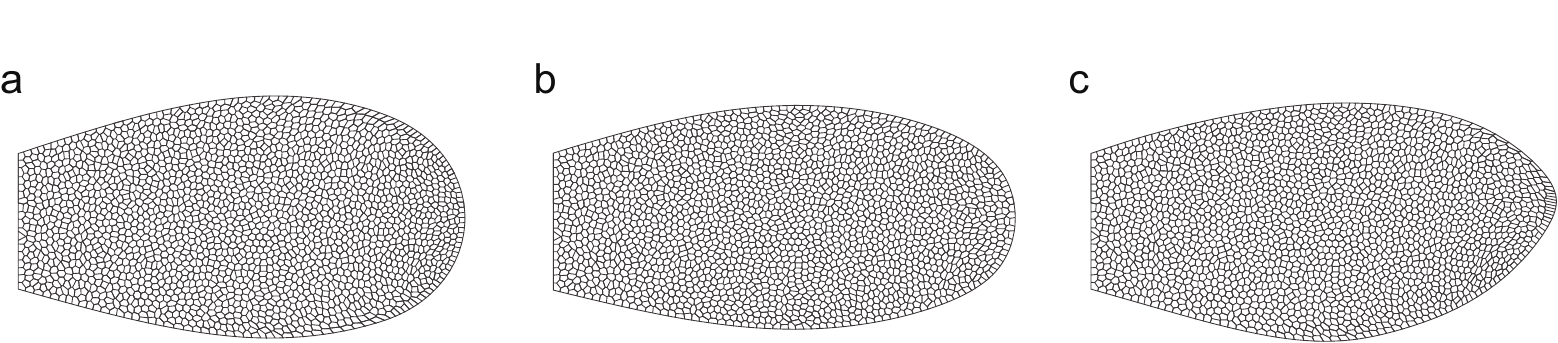}
  \caption{Simulation results of the integrated mechanical model of the wing blade and margin for different distributions of the bending stiffness.
  (a) A result with uniform bending stiffness showing a curve similar to Fig. \ref{fig:elastica}(b).
  (b) A result with an anterior-posteriorly symmetric distribution of the bending stiffness as in Fig. \ref{fig:nonuniformK-Simple}(a2). Smaller values in the distal region change the wing shape, notablly in the distal region.  
  (c) A result with an anterior-posteriorly asymmetric distribution of the bending stiffness as in Fig. \ref{fig:nonuniformK-Simple}(b2). A bulge in the posterior region
is formed. 
 All results contain 2107 homogeneous cells and the parameters of the cell vertex model in Fig. 10 of \cite{Ishimoto14} is used. Each result was obtained, at least, after $4,000,000$ time steps with adaptive time stepping. In order to realise the pinned boundary condition, the vertices on the proximal side of the wing blade are constrained on the same line while the connecting angles of the margin to the side are not.
   }
  \label{fig:hybrid}
\end{figure*}

\section{Discussions}
\subsection{Implication on the evolutionarily conserved mechanism for the determination of insect wing margin shapes}
\label{sec:summary}
Motivated by the beauty of smooth curves of insect wings,
we have proposed and constructed a mechanical model for the diversified insect wing margin shapes by generalising the Euler's elastica.
It is assumed that the wing margin has bending stiffness as its multicellular mechanical properties, and
our numerical simulations showed that the spatial distribution of bending stiffness of the wing margin was sufficient
to reproduce diverse wing margin shapes found in natural insects.
Although identification of such non-uniform distribution of the bending stiffness in a developing wing tissue awaits future studies,
the observation that proximal anterior, distal anterior, and posterior domains of the wing margin contain a unique set of differentiated cells
\cite{Garcia-Bellido1972, Couso1994} 
suggest that the domains might have distinctive mechanical properties and genes encoding mechanical structural/signaling components might be differentially expressed along the margin contour.
Interestingly, the inferred profiles of the bending stiffness were distinct for different insect images,
but all shared spatial (proximal anterior, distal anterior, and posterior domain) domain structure.
Thus, similar to the margin cell differentiation \cite{Garcia-Bellido1972}, the margin shape determination might also be under the control of global patterning of the wing.
Taken all the above into consideration, we speculate that diverse wing margin shapes in different species might have evolved by modifying 
the conserved, mechanical machineries at the downstream of patterning information.

\subsection{Notes on the spatial profile of the bending stiffness of the wing margin}
\label{sec:kappasignificance}
One might argue that the complexity of the inferred bending stiffness function exceeds by far the complexity of the gene patterning along the margin.
However, several lines of evidence imply that this may not be the case.
First, inferred bending stiffness takes collective values around the margin, absorbing relevant surrounding mechanical properties.
For instance, we found that the complex profile of inferred $\kappa(s)$ was often associated with the position of veins,
which may pull the margin towards the hinge as suggested in a previous study \cite{Ray2015}.
Second, the inferred bending stiffness in Fig. \ref{fig:infer} appears 
to be complex because it contains noise generated by image acquisition and processing. 
Third, as shown in Fig. \ref{fig:nonuniformK-Simple} in Sec. \ref{sec:simulation},
a simple distribution of the bending stiffness is sufficient to reproduce a wing-like shape in a numerical simulation of our model.
From these arguments and results, we speculate that the spatial distribution of the bending stiffness in real insects may be sufficiently simple and smooth
such that it can be coded by global patterning of the wing.

A measurement of the spatio-temporal profiles of bending stiffness of the wing margin is expected to directly prove the hypothesis of our model.
Because, in the case of a beam, the bending stiffness can be estimated by the Young's modulus multiplied by the second moment of area,
corresponding quantities of the wing margin are to be measured.
Once the bending stiffness of the wing margin is measured, one can quantitatively assess the relationship between 
the spatial variation of the bending stiffness and the wing shape by genetically or pharmacologically manipulating the bending stiffness.

\subsection{Future extensions of the current modelling framework}
There are several possibilities to extend the current framework of modelling. 
First, additional material properties of the wing margin such as extensibility can be considered.
Second, future work will study how the wing acquires its margin shape during development.
For this, one needs to consider temporal changes in the bending stiffness profiles and/or morphogenetic dynamics of the margin cells
(\emph{i.e.}, cell shape change, cell rearrangement, cell division, and apoptosis) \cite{Baonza2000, Takemura2011},
which can possibly couple to the extensibility of the wing margin as well.
Third, the mechanical interaction between the wing margin and other structural components of the wing is definitely a research direction to pursue.
For instance, it has been shown that patterned linkage between the wing epithelial cells and the overlying chitinous cuticle
is essential to give rise to the tissue tension that shapes the \textit{Drosophila} wing \cite{Etournay2015, Ray2015}.
Also, the inferred profiles in Fig. \ref{fig:infer}(f) may be influenced to some extent by the positions of the margin connected to the veins,
as some characteristic points in the profiles correspond to such positions. 
Though the above observations have not clearly been described in the language of mechanics,
the next direction of study is to merge our model of the margin with other mechanical structures either in a discrete or in a continuous way \cite{Khalilgharibi2016}.
One of the discrete candidates for the replacement of our static wing blade is the cell vertex model [46--49],
as we presented in our first attempt in this direction in Sec. \ref{sec:IntegrateModel} (Fig. \ref{fig:hybrid}).
The existence of the veins can be realised on the hybrid model by introducing specific cell types
such as vein cells or some specific mechanical properties to the borders between intervein and margin cells.
For instance, a recent study has shown that increasing the line tension at the vein-intervein boundaries
is required to reproduce the wing shape \cite{Ray2015}.
A continuous candidate is a continuum body equipped with some appropriate constitutive equations to be constructed and to be consistent with the existing experimental observations \cite{Etournay2015, Guirao2015, Sugimura2016}.
Both approaches would provide a better understanding of the wing shaping and more refined tools to compare or support experimental observations with computational modelling.

\subsection{Mathematical notes and comments on the models}
\label{sec:mathnotes}
For the simulations in Fig. \ref{fig:nonuniformK}, 
we have always found an appropriate $\varphi(0)$ for the pinned-pinned boundary condition for any tested form of $\kappa(s)$. 
Although we have not shown this rigorously, it suggests another  
unravelled law of conservation analogous to the energy conservation law in the Euler's elastica. 
This is a mathematically interesting statement to prove.

As for the stiffness inference, the dependence of the inference on the initial value of $\kappa(0)$ has not been investigated.
The value cannot be scaled out of Eq. (\ref{eq:diff}), so that its investigation might be necessary in some cases,
particularly when the wing margin buckles due to the internal pressure.
In such a case, there would exist singular points such as inflection points of $\varphi(s)$, and their positions may strongly depend on the initial value.
In addition, in the search for buckled shapes in Fig. \ref{fig:extended}, we have sometimes found singularities at the end points.
This suggests that if the wing margin contains a buckled part, for example in its earlier developmental stages,
special care is required for the stiffness inference.
This is out of the scope of the present study, but is surely an interesting point to investigate further.

We have shown that internal pressure/tension affects both on the shape and inferred bending stiffness of the wing margin.
While the internal pressure/tension provides additional types of wing margin shapes obtained by simulation as shown in Fig. \ref{fig:extended},
the allowed value of the negative pressure (\emph{i.e.}, tension) was found to be very restrictive because of the physical requirement, $\kappa(s)\geq 0$. 
This restriction may be relaxed by considering additional force generators, such as veins, which can restore the stiffness back to the physical region. 
This is to be investigated further in conjunction with the above mentioned hybrid models.

\subsection{Conclusion}
The present study provides the simple yet insightful approach
for understanding the mechanical control of the insect wing margin shape. 
We expect that it will serve as a basic building block of an integrated model of wing development in the future.

\vspace*{30pt}
\noindent 
{\bf Acknowledgments}

The authors are grateful to Yoshihiro Morishita, Alexis Matamoro-Vidal,
Fran\c{c}ois Graner, Philippe Marcq, Daiki Umetsu, Tsuyoshi Hirashima, Frank J\"ulicher, 
and Osamu Shimmi for their stimulating discussions and suggestions.
We would like to thank Stephanie Nix and the other members of the laboratory for Life-Integrated Fluid Engineering 
in Akita Prefectural University for their discussions.
We would also like to acknowledge the current and past members of the laboratory 
for Developmental Morphogeometry in RIKEN QBiC for their continuous support and stimulating discussions.
The present work was commenced while YI was in the laboratory. 
This work was supported by JSPS KAKENHI (Grants-in-Aid for Scientific Research) Grant Number 26540158 and 17K00410 to YI and 
by JST PRESTO (JPMJPR13A4) to KS.

\newpage
\appendix
\vspace{20pt} \noindent {\Large\bf Appendix}

\section{Derivation of the general solution of the Euler's elastica}
\label{app:deriv}

In this section, we show the derivation of the general solution of the Euler's elastica (Eq. (\ref{eq:Euler})). 
\bea
    \kappa  \ddot \varphi(s)  = -F \sin \varphi(s)
\eea
Here, we use the shorthand notation of the $s$-derivative: $\dot\varphi\equiv\frac{d\varphi}{ds}$.
Multiplying both sides of the equation by $\dot \varphi(s)$ and integrating over $s$ from $0$ to $L$ leads
\bea
  \frac{\kappa}{F} \int_0^L ds' \dot\varphi(s') \ddot\varphi(s') &=& - \int_0^L ds' \dot\varphi(s')  \sin \varphi(s')
\nn
  \frac{\kappa}{2F} \left( \dot\varphi(s) \right)^2
  &=&  \frac{\kappa}{2F} \dot\varphi(0)^2 + \cos \varphi(s) - \cos\varphi(0)
\nn
  \dot\varphi(s) &=& \pm \sqrt{ \frac{2F}{\kappa}\left( \cos\varphi(s) - \cos\varphi(0) \right) + \dot\varphi(0)^2  }
\nn
  ds &=& \pm \frac{d\varphi(s)}{A \sqrt{\cos\varphi(s)-B}},
\eea
where $A\equiv\sqrt{\frac{2F}{\kappa}}$ and $B\equiv\cos\varphi(0)-\frac{\kappa}{2F}\dot\varphi(0)^2$.
The sign on the right hand side depends on the direction in which $\varphi(s)$ changes.
When $\varphi(s)$ decreases as $s$ increases, the sign is negative. 
In the context of pendulum, $B$ is the energy transferred to the kinetic energy at $s$.
By the symmetry argument, $\varphi(s)$ reaches zero at $s=L/2$.
Integrating the above again over $s$ from $0$ to $L/2$, one obtains the following solution
\bea
\frac{L}{2} &=& \int_{0}^{\varphi(0)} \frac{d\varphi}{A \sqrt{\cos\varphi-B}}
\nn
  &=& \frac{1}{A\sqrt{1-B}} \int_{0}^{\varphi(0)} \frac{d\varphi}{\sqrt{ 1 - \frac{2}{1-B}\sin^2\varphi/2}}
\nn
  &=& \frac{2}{A\sqrt{1-B}}\, F \left( \left. \frac{\varphi(0)}{2} \right\vert \frac{2}{1-B} \right), 
\eea
where $F(z|m)$ is the incomplete elliptic integral of the first kind.
This can be further simplified when $\varphi(0)$ is a multiple of $\pi$: $F(k\pi/2|m) = k K(m)$,
where $K(m)$ is the complete elliptic integral of the first kind.
The above solution is the relation between $L$, $\kappa$, and $F$.
Thus if the initial conditions and $L$ are given, the ratio $\kappa/F$ is also given.

\section{Units and nondimensionalisation}
\label{sec:nondimensionalization}

The units of the relevant quantities are
\bea
\left[ s \right] &=& \left[ L \right] , \nn
\left[ \kappa(s) \right] &=& [M][L]^3 [S]^{-2} , \nn 
\left[ F \right] &=& [M][L][S]^{-2} ,\nn 
\left[ p \right] &=& [M][S]^{-2} ,\nn
\left[ \varphi(s) \right]   &=& [M]^0 ,
\eea
where $[M], [L]$, and $[S]$ stand for the dimensions of mass, length, and time, respectively.
Thus $\frac{\kappa(s)}{F}$ and $\frac{p}{F}$ have the dimensions $[L]^2$ and $[L]^{-1}$.

Redefining $\kappa(s)/F$ and $p/F$ by $\kappa(s)$ and $p$, respectively, and normalising $s$ by the total length of the margin $L$,
we have dimensionless combinations of $\kappa(s)/L^2$ and $p L$. 
With the dimensionless variable and parameters
$\wt s \equiv s/L$, $\wt \kappa(\wt s) \equiv \kappa(s)/L^2$, $\wt p \equiv p L$, and $\wt\varphi(\wt s)\equiv \varphi(s)$,
Eqs. (\ref{eq:Euler},\ref{eq:nonuniformK},\ref{eq:p}) can be nondimensionalised as
\bea
  \label{eq:nondimensionalised set}
  \wt \kappa \frac{d^2 \wt\varphi(\wt s)}{d \wt s^2} &=& - \sin \wt\varphi(\wt s),
\nn
  \wt \kappa(\wt s) \frac{d^2 \wt\varphi(\wt s)}{d \wt s^2} + \frac{d \wt\kappa(\wt s)}{d \wt s} \frac{d \wt\varphi(\wt s)}{d \wt s} &=& - \sin \wt\varphi(\wt s),
\nn
  \frac{d}{d \wt s} \left( \wt\kappa(\wt s) \frac{d \wt\varphi(\wt s)}{d \wt s} \right)
  &=& -\sin \wt\varphi(\wt s) + p \Vec{u}(\wt s) \cdot \left( \wt{\Vec{r}}(\wt s) - \frac{\wt{\Vec{r}}(1)}{2}  \right).
\eea
These tilded quantities are implied in the manuscript when the above nondimensionalisation is mentioned.

\section{Derivation of the deterministic equation of the homogeneous pressure model}
\label{sec:derivation_extended}

In this section, we show how to derive Eq. (\ref{eq:p}) of the homogeneous pressure model in Sec. \ref{sec:extended}.
There are two ways to formulate the model.
Of the two, we employed the principle of virtual work, or of the least action,
because the force balance formulation cannot be given in a naive way with the internal pressure.

Following the standard way of formulating the elastic medium \cite{landau}, one can write the energy function of the wing margin as
\bea
  E_{bend} &=& \int_0^L ds \frac{\kappa(s)}{2} \left( \frac{d\varphi(s)}{ds} \right)^2.
\eea
Then, one can equate the variation of the energy with the variation of the virtual work done by the external forces as
\bea
  \delta E_{bend} &=& \delta W, 
\eea
where 
\bea
  W &=& F \left( L-|\Vec{r}(L)| \right) + p A 
\nn
  &\simeq& - F\int_0^L ds \cos\varphi(s) + p A.
\eea
Here, $F$ is the pinching force acting on the ends of the wing margin.
Going from the first line to the second, $r_y(L)=0$ is implicated and the term $FL$ is omitted
since it does not contribute to the variation of the work.
$p$ is the internal pressure, or the tension when it is negative, while $A= \frac12 \int_0^L ds (\Vec{u}(s)\times \Vec{r}(s))$ is the area of the wing blade.
In other words, the hinge plays a role of the reservoir for the pressure and keeps supplying the pressure $p$ through the proximal boundary of the wing blade. 

The corresponding action can be expressed by $S\equiv E-W$:
\bea
  S &=& \int_0^L ds \left[ \frac{\kappa(s)}{2} \left( \frac{d\varphi(s)}{ds} \right)^2 
   + F \cos\varphi(s) - \frac{p}2 \,\Vec{u}(s)\times \Vec{r}(s)
 \right].
\eea
The variation of the last term by $\delta\varphi$ can be given by
\bea
  \delta\left( \int_0^L ds\, \Vec{u}(s)\times \Vec{r}(s) \right)
&=& \delta\left\{ \int_0^L ds\, \Vec{u}(s) \times 
   \left( \int_0^s dt \cos\varphi(t), \int_0^s dt \sin\varphi(t) \right)^T \right\}
\nn
  &=& \int_0^L ds\, \delta\varphi(s) \left\{ (-\sin\varphi(s),\cos\varphi(s))^T \times
    \Vec{r}(s) \right\}
\nn&&
  + \int_0^L dt\, \delta\varphi(t) \left\{ 
    \left( \int_t^L ds \Vec{u}(s) \right) \times \left(  
      -\sin\varphi(t), \cos\varphi(t)
    \right)^T
    \right\}
\nn
  &=& \int_0^L ds\, \delta\varphi(s) \left\{ 
     - \Vec{u}(s)\cdot\Vec{r}(s)
     + (\Vec{r}(L)-\Vec{r}(s) ) \cdot \Vec{u}(s)
  \right\}
\nn
  &=& - 2 \int_0^L ds\, \delta\varphi(s) \,
     \Vec{u}(s) \cdot \left( \Vec{r}(s) - \frac{\Vec{r}(L)}{2} \right).
\eea
Note, for the derivation of the second term in the second line, we have changed the regions of integration over $s$ and $t$ by the identity
\bea
  \int_0^L dx \int_0^x dy\, f(x,y) = \int_0^L dy \int_y^L dx\, f(x,y) .
\eea
By applying the least action principle $\delta S=0$, one finds Eq. (\ref{eq:p}) as
\bea
  0 &=& -\frac{d}{ds}\left( \kappa(s) \frac{d\varphi(s)}{ds} \right)
     - F \sin\varphi(s) + p\, \Vec{u}(s)\cdot \left( \Vec{r}(s)-\frac{\Vec{r}(L)}{2} \right) .
\eea

\section{Derivation of the difference equation for the stiffness inference}
\label{sec:inference}

Let us start with nondimensionalised Eq. (\ref{eq:nonuniformK}) with the dots
\bea
\label{eq:original}
  \kappa(s) \ddot\varphi(s) + \dot\kappa(s) \dot\varphi(s) = - \sin\varphi(s).
\eea
Given $\ddot\varphi(s_n)$, $\dot\varphi(s_n)$, $\varphi(s_n)$, $\kappa(s_{n-1})$, and $\dot\kappa(s_{n-1})$,
where $\{s_n\}$ is the set of discrete points from an image, we sought to infer the values of $\dot\kappa(s_n)$ and $\kappa(s_n)$.
If the two-dimensional position is defined at $s_n$ by $\Vec{s}_{n}$ and $\Delta s_n \equiv |\Vec{s}_{n+1} - \Vec{s}_{n}|$,
then $\kappa(s_n)$ can be given exactly by the form of the forward difference at $s_{n-1}$ by
\bea
\label{eq:kappa}
  \kappa(s_n) &=& \kappa(s_{n-1}) + \Delta s_{n-1} \delta_{f} \kappa(s_{n-1}),
\nn
  \delta_{f} \kappa(s_{n-1}) &=& \frac{\kappa(s_n) - \kappa(s_{n-1})}{\Delta s_{n-1}}.
\eea
This forward difference $\delta_f$ at $s_{n-1}$ is equivalent to the central difference $\delta_c$ at $s_{n-\frac12}$, which is the middle point between $s_{n-1}$ and $s_n$. Assuming $\Delta s_{n-1} \ll 1$, we approximate this central difference by the average of the first derivatives at $s_{n-1}$ and $s_n$ as
\bea
  \delta_{c} \kappa \left(s_{n-\frac12}\right) &=& \frac{\kappa(s_{n}) - \kappa(s_{n-1})}{\Delta s_{n-1}}
\nn
   &\simeq& \frac{\dot\kappa(s_{n-1}) + \dot\kappa(s_n)}{2}.
\eea
This approximation leads to the second line of Eq. (\ref{eq:diff}).
Plugging this into the expression (\ref{eq:kappa}) and to Eq. (\ref{eq:original}), one gets
\bea
  \dot \kappa(s_n) \left( \dot \varphi(s_n) + \frac{\Delta s_{n-1}}2 \ddot \varphi(s_n) \right)
  + \left( \kappa(s_{n-1}) + \frac{\Delta s_{n-1}}2 \dot \kappa(s_{n-1}) \right) 
    \ddot \varphi(s) = - \sin\varphi(s).
\eea
By replacing $s_{n-1}, s_{n}$ and $\Delta s_{n-1}$ by $s-\Delta s$, $s$ and $\Delta s$ for simplicity, the above can be expressed by Eq. (\ref{eq:diff}). Similarly, the difference equation for the homogeneous pressure model can be given trivially .

We have derived and used the difference equation (\ref{eq:diff}) for the stiffness inference. 
There could be a variety of ways to make the differential equations the difference ones, which are out of the scope of the current manuscript.


\end{document}